\documentclass[11pt]{article} 
\newcommand{\RRR}{$I\hspace{-0.25em}R^3$}

\newcommand{\bea}{\begin{eqnarray}} 
\newcommand{\eea}{\end{eqnarray}} 
\newcommand{\be}{\begin{equation}} 
\newcommand{\ee}{\end{equation}} 
\newcommand{\bdm}{\begin{displaymath}} 
\newcommand{\edm}{\end{displaymath}} 
\newcommand{\bq}{\begin{quote}} 
\newcommand{\eq}{\end{quote}}

\begin{document}
\setlength{\baselineskip}{16pt}
\title{A SPACE FOR THE QUANTUM WORLD}
\author{Ulrich Mohrhoff\\ 
Sri Aurobindo International Centre of Education\\ 
Pondicherry-605002 India\\ 
\normalsize\tt ujm@satyam.net.in} 
\date{}
\maketitle 
\begin{abstract}
\normalsize\noindent
Epistemic interpretations of quantum mechanics fail to address the puzzle 
posed by the occurrence of probabilities in a fundamental physical theory. 
This is a puzzle about the physical world, not a puzzle about our relation to 
the physical world. Its solution requires a new concept of physical space, 
presented in this article. An examination of how the mind and the brain 
construct the phenomenal world reveals the psychological and 
neurobiological reasons why we think about space in ways that are 
inadequate to the physical world. The resulting notion that space is an 
intrinsically partitioned expanse has up to now stood in the way of a 
consistent ontological interpretation.
\setlength{\baselineskip}{14pt}
\end{abstract}

\section{\large INTRODUCTION}

This article presents a new concept of physical space. Section~2 shows 
that the behavior of electrons in two-slit experiments is inconsistent with the 
notion that physical space is an intrinsically partitioned expanse. Section~3 
explains why epistemic interpretations of quantum mechanics (QM) fail to 
address the puzzle posed by the occurrence of probabilities in a 
fundamental physical theory. This is a puzzle about the physical world, not 
a puzzle about our relation to the physical world. However, it is by 
examining how the mind and the brain construct the {\it phenomenal} world 
that we come to understand why we are prone to think about space in ways 
that are inadequate to the {\it physical} world. The concept of an intrinsically 
partitioned space stands in the way of a consistent {\it ontological} 
interpretation of QM. In epistemic interpretations the continuous extension 
and the spatial multiplicity of the world are phenomenologically grounded. 
In an ontological interpretation these features have to be understood as 
aspects of a strongly objective world. How this may be done is discussed in 
Sec.~4. Section~5 addresses the relation of the theory to the actual 
physical world. The key to this core interpretational issue is the objective 
indefiniteness that contributes to ``fluff out'' matter. It entails the extrinsic 
nature of the values of quantum-mechanical observables, which implies the 
contingent reality of spatial distinctions and the finite spatial differentiation 
of the physical world. Owing to the latter, the field-theoretic notion of an 
intrinsically and infinitely differentiated space is as inconsistent with QM as 
the notion of absolute simultaneity is with special relativity. Section~6 
contains concluding remarks.

\section{\large THE TWO-SLIT EXPERIMENT REVISITED}
\label{TSE}
The strange behavior of electrons in two-slit 
experiments~\cite{Feynmanetal65} has led David Albert to the conclusion 
that ``[e]lectrons seem to have modes of being, or modes of moving, 
available to them which are quite unlike what we know how to think 
about''~\cite{Albert92}. Albert arrives at this conclusion after considering 
what he claims to be ``all of the logical possibilities that we have any notion 
whatever of how to entertain.'' Each of the considered possibilities finds 
expression in a conjunction of two propositions, and thus is equivalent to 
the affirmation of two truth values, one for the proposition "The electron 
went through the left slit" (symbolically, $e\rightarrow L$, or $\cal L$ for 
short) and one for the proposition "The electron went through the right slit" 
($e\rightarrow R$ or $\cal R$):

\medskip(1) $\quad e\rightarrow L$ \& $e\not\rightarrow R$,\par
(2) $\quad e\not\rightarrow L$ \& $e\rightarrow R$,\par
(3) $\quad e\not\rightarrow L$ \& $e\not\rightarrow R$,\par
(4) $\quad e\rightarrow L$ \& $e\rightarrow R$.

\medskip\noindent 
In the quantum domain, the sufficient and necessary condition for the 
existence of a truth value is a fact (an actual event or an actual state of 
affairs) from which a truth value can be 
inferred)~\cite{Mohrhoff00,UMABL,UMUR,UMWATQM}. If nothing 
indicates the truth value of $\cal L$ then $\cal L$ is neither true nor false 
but meaningless. In this case neither the property ``through $L$" 
(represented by a subspace $\bf L$ of a Hilbert space $\cal H$) nor the 
property ``through $R$" (represented by the orthocomplement of $\bf L$ in 
$\cal H$) can be attributed to the electron, nor does the corresponding 
binary observable (represented by the projector onto $\bf L$) have a value. 
{\it No property is a possessed property unless it is an indicated property.} 
In other words, the values of quantum-mechanical observables are {\it 
extrinsic} (possessed because they are indicated) rather than {\it intrinsic} 
(indicated because they are possessed). They supervene on 
property-indicating facts.

If any of the above four conjunctions is true, no interference is observed. 
(The fourth conjunction is of course never true.) Albert's conclusion that 
electrons can move in ways that we do not know how to think about is 
nevertheless unfounded, for his list is incomplete. There is another logical 
possibility:

\medskip (5) $\quad e\rightarrow L\& R$.

\medskip\noindent $L\& R$ stands for the two slits considered as a whole. 
If the electron originates in front of the slit plate, and if $L$ and $R$ are the 
only openings in the slit plate, then (5) is true if and only if the electron is 
detected behind the slit plate. As the following section will show, the reason 
we tend to miss this possibility lies in our neurobiological make-up. This 
predisposes us to conceive of space as an intrinsically partitioned 
expanse---to believe that, no matter how we conceptually partition a spatial region, 
the individual partitions have an objective reality and are objectively 
distinct. This leads to the notion that space is intrinsically (``by itself'') 
partitioned into infinitesimal regions, and this notion, if it were correct, 
would justify the customary mathematical representation of space by the 
set \RRR\ of triplets of real numbers. But if every region of space were 
intrinsically and infinitely partitioned, proposition~(5) would imply that either 
proposition~(1) or proposition~(2) is true. If $L$ and $R$ exist by 
themselves as distinct ``parts of space'' then nothing goes through $L\& R$ 
without going through either $L$ or $R$ {\it and} without being divided into 
parts that go through different slits---the distinctness of the regions would 
imply the existence of distinct parts.

What the interference fringes are trying to tell us is that an electron is quite 
capable of going through $L\& R$ without going through either $L$ or $R$ and without 
being divided into parts. Proposition (5) does not imply that either 
proposition~(1) or proposition~(2) is true. The reason this is so is the 
extrinsic nature of possessed positions. Only indicated positions are 
possessed, and the only indicated positions are the finite sensitive regions 
of detectors. If the truth of proposition~(5) is indicated then proposition~(5) 
is true. If at the same time nothing indicates truth values for the propositions 
$\cal L$ and $\cal R$ then the electron's position at the time of its passing 
the slit plate is $L\& R$, and it is not any smaller region inside $L\& R$. 
The individual regions $L$ and $R$ do not exist for the electron, and 
therefore they cannot ``force the electron to chose between them.'' A 
region~$V$ exists for an object~$O$ at a time~$t$ if and only if the 
proposition ($\cal P$) ``$O$~is in~$V$ at~$t$'' has an (indicated) truth 
value. Space is not an intrinsically partitioned expanse. Partitions of space 
are {\it contingent}. A partition may be real for one object and not for 
another, at one time and not at another, depending on which propositions of 
the form $\cal P$ possess (indicated) truth values.

Most readers will have seen pictures of hydrogen orbitals. As is well known, 
the ``electron cloud'' represents a probability density 
$|\psi_{nlm}(r,\theta,\phi)|^2$ rather than some ``kind of bizarre real 
jelly''~\cite{dE79}. It is impossible to form a realistic image of an atom. The 
reason this is so is that nothing in the physical world corresponds to the 
intrinsically differentiated canvas of our mental images. We cannot help 
perceiving, and conceiving of, the space over which the electron cloud is 
extended as being divided into mutually disjoint regions. We may not 
actually imagine any particular partition of that space, but the idea that 
disjoint regions are individually existing and mutually distinct ``parts of 
space'' underlies our theoretical dealings with the 
world~\cite{UMCCP,BCCP}.

If $|\psi_{nlm}(r,\theta,\phi)|^2$ is the right density for assigning 
probabilities to regions in the space of the hydrogen atom's internal relative 
position~$\bf R$, these regions exist solely in our imagination; they have 
no counterparts in the physical world. For this reason all we can say about 
$\bf R$ is counterfactual and probabilistic~\cite{Mohrhoff00,UMABL}. 
Although we know very well that no physical detector (consisting 
necessarily of a large number of atoms) can monitor a region of space 
smaller than the size of an individual atom, we need to assume the 
existence of an array of detectors monitoring a set $\{R_i\}$ of such 
regions, and we need to assume that exactly one detector clicks. If these 
conditions are fulfilled (which they never are) the integral of $|\psi|^2$ over 
each region $R_i$ gives the probability that the corresponding detector is 
the one that clicks. The detectors are needed to realize (make real) the 
mutually disjoint regions, if only counterfactually, for a region $R_i$ is real 
if and only if a truth value exists for the proposition ``The value of $\bf R$ is 
in~$R_i$.'' (To be precise, the existence of a truth value is equivalent to the 
reality of $R_i$ {\it for} $\bf R$. But if $R_i$ is not real for the electron's 
position relative to the proton, it is not likely to be real for the position of {\it 
any} object relative to the proton. And if a region is not real for any object, it 
is not real at all.)

\section{\large THE MIND, THE BRAIN, AND THE WORLD}

As early as 1923, the year in which Louis de Broglie introduced electron 
waves and explained the quantization of orbital angular momentum 
postulated by Niels Bohr, Bohr wrote, ``It is my personal opinion that these 
difficulties are of such a nature that they hardly allow us to hope that we 
shall be able, within the world of the atom, to carry through a description in 
space and time that corresponds to our ordinary sensory 
perceptions"~\cite{Bohr23}. Bohr did not say that a spatiotemporal 
description is impossible but only that such a description cannot be 
modeled after our ordinary sensory perceptions. This distinction seems to 
have been lost on his contemporaries, who instead of looking for an 
adequate spatiotemporal description discovered or 
re-invented Kant's theory of science. Like Kant, they believed that space 
and time lie in the mind of the beholder. Science, accordingly, does not 
deal with the unperceived world; it deals with how the world appears to us, 
on the 3+1-dimensional canvas of mental space and time. Hence their 
insistence on the epistemic nature of 
QM~\cite{LonBau,vN,Wigner,Heisbg,Peierls}.

As I see it, the key to a major puzzle posed by QM lies in the careful 
distinction between phenomenal space and physical space. Phenomenal 
space is the spatial aspect of the phenomenal world, of which we are 
directly aware. Physical space is the spatial aspect of the physical world 
(the world investigated by physical science) whose properties we need to 
discover or infer since we are not directly aware of them. The Kantian 
stance is appropriate for dealing with the phenomenal world. It is 
appropriate for classical physics because classical physics is consistent 
with the Kantian idea that the objective world is the totality of what appears 
(on the mental canvas of space and time). It is inappropriate for quantum 
physics because quantum physics is inconsistent with this idea, inasmuch 
as electrons never appear on the mental canvas.

Supporters of epistemic interpretations might counter that this is so 
because there are no electrons. According to Ole Ulfbeck and Aage 
Bohr~\cite{UlfBohr}, there are no electrons or neutrons; there are only 
electron clicks and neutron clicks. This point of view is founded on the 
claim that space and time have no mind-independent existence. If electrons 
do not appear (on our mental canvas), and if spatiotemporal concepts 
cannot be applied to what does not appear, then we can have no 
knowledge of the properties of electrons. And if we cannot know their 
properties, we cannot know that they exist. Yet we do---how else could we 
talk about electron clicks and neutron clicks?---and for this reason we must 
reject the claim that spatiotemporal concepts are applicable only to what 
appears.

The antithetical claim has also been made that QM affords us a glimpse of 
the ``veiled reality''~\cite{dEVR} beyond the domain of experienced facts, 
and that this compels us to acknowledge something like the Kantian 
dichotomy between the world-as-we-know-it and the world-in-itself, and 
hence the merely intersubjective (weakly objective) reality of facts. 
However, as Kant's idealistic successors were quick to point out, a reality 
that is by definition beyond our ken can play no role in our theoretical 
accounts of the world. Whereof we cannot say what it is thereof we must 
remain silent. Contrariwise, if we can meaningfully speak of what lies 
behind the ``veil of appearances,'' this forms part of the empirically known 
world. There is a dichotomy that is essential to QM, namely the dichotomy 
between a ``classical domain'' of property-indicating facts and a ``quantum 
domain'' of properties indicated by facts, but this is a dichotomy {\it within} 
the empirically known world. It has nothing to do with the Kantian 
dichotomy~\cite{Mohrhoff00,UMUR}. Only if it is decided beforehand that 
there are no spatiotemporal concepts appropriate to the quantum domain, 
does the quantum domain become unknowable like Kant's world-in-itself.

Under the influence of quantum information theory something like the 
Kantian dichotomy is nonetheless gaining renewed popularity. (See, for 
instance, the correspondence of Christopher Fuchs with several 
distinguished physicists~\cite{Fuchs}.) The idea seems to be that QM is 
necessarily about information {\it because} the information it is concerned 
with is about an unknowable reality. Quantum information, however, is not 
about anything unknowable. It is about possibilities and conditional 
probabilities: On condition that the values of a given set of observables 
$Q_1,\dots,Q_n$ are indicated at the given times $t_1,\dots,t_n$, QM 
assigns a joint probability to each set of possible values, on the basis of 
some chosen set of relevant facts~\cite{UMABL}. It is therefore 
gratuitous to portray QM as being fundamentally about information. Since 
quantum information concerns probabilities, QM concerns probabilities {\it 
rather than} information.

The puzzle posed by the occurrence of probabilities in a fundamental 
physical theory is solved neither by invoking a reality beyond our ken nor 
by invoking the subjective aspect of our existence. It is a puzzle about the 
empirically known world, not a puzzle about our relation as conscious 
subjects either to the known world or to whatever lies beyond 
it~\cite{Mohrhoff00,UMUR,UMWATQM,MohrhoffFF}. However, while such 
words as ``consciousness,'' ``experience,'' or ``information'' have no 
legitimate place in interpretations of QM, an examination of the processes 
by which the mind and the brain co-produce the {\it phenomenal\/} world, 
can take us a long way toward understanding why we find it so hard to 
make sense of the {\it physical\/} world.

Consider the disparate ways in which we think about (i)~a position or 
region of space and (ii)~any other feature of the phenomenal world. We 
readily agree that the color of a ripe tomato or the shape of a sphere 
cannot exist by itself (not, at least, in the material world) without the 
existence of a material object of which it is the color or the shape. Yet we 
continually behave (mentally) as if a position or a region of space were 
something that exists by itself, whether or not there is a material object to 
which it can be attributed. What is ultimately responsible for the 
disparateness between our (conceptual) handling of positional information 
and our handling of other sensory data is the process by which the 
mind/brain system integrates into phenomenal objects such phenomenal 
variables as hue, lightness, shape, and motion. This integration is based on 
positional information. Phenomenal variables that occur in the same place 
are perceived as features of the same object, while phenomenal variables 
that occur in different places are perceived as features of different objects. 
Positional information thus plays a unique role in the process of feature 
integration. This uniqueness is reflected in the functional organization of 
the brain, where feature maps seem to be everywhere. (A feature map is a 
layer of the neocortex in which cells map a particular phenomenal variable 
in such a way that adjacent cells generally correspond to adjacent locations 
in the visual field. In the macaque monkey as many as 32 distinct visual 
feature maps have been identified.) While every phenomenal variable 
except location has at least one separate map, locations are present in all 
maps as the integrating factors~\cite{Clark}.

While the features of each perceived object share the same phenomenal 
location, their neural correlates are scattered across a good many locations 
in the neocortex, and these different physical locations do not project (via 
neural pathways) on any single physical location: The brain lacks an 
``object map.'' There is a master map where ``it all comes together,'' but 
this is not physical; it is the spatial aspect of our mental 
canvas---phenomenal space. This suggests to me that Kant was right in his insistence 
that space is primarily a subjective expanse, and that, epistemologically 
speaking, physical space is a projection or objectification of this subjective 
expanse. This, however, leaves open the question as to whether physical 
space is {\it nothing but} such a projection, or whether it is a 
mind-independent, strongly objective expanse, of which this subjective expanse 
is a more or less faithful reproduction. We ought not to anticipate the 
answer to this question by assuming beforehand that physical space is 
nothing but such a projection. Instead we should allow for the strong 
objectivity of physical space and deduce from this its actual properties. The 
question to be addressed, then, is this: Which of the properties of 
phenomenal space can be attributed to (a strongly objective) physical 
space, under which conditions, and to what extent? The answer to this 
question is not an all-or-nothing affair, as those implicitly assume who 
conclude that only the classical domain admits of a spatiotemporal 
description.

Recall the special role played by positional information in the construction 
(by the mind/brain system) of the phenomenal world. If we accord a 
mind-independent reality to the physical world, positional information cannot play 
an analogous role in the creation of the physical world. (If there are no 
neural correlates of the {\it physical\/} world, its creation cannot involve 
anything like neural feature maps.) The expanse of {\it phenomenal\/} 
space and a multiplicity of distinct {\it phenomenal\/} locations exist 
independently of what appears, for this expanse is a feature of our mental 
canvas, and the multiplicity of locations is rooted in our neuroanatomy---in 
the distinct physical locations of the neurons in each feature map. There is 
no corresponding foundation for the independent existence of {\it physical\/} 
space---independent of material 
objects---nor for the independent existence of a multiplicity of {\it physical\/} 
locations. The belief in the independent existence of an intrinsically 
differentiated physical space is therefore unwarranted. Not only is it 
unwarranted but it also prevents us from understanding QM. The fact that 
this belief is, as it were, ``hard-wired,'' explains why we find it so hard to 
beat sense into QM.

\section{\large THE NATURE OF PHYSICAL SPACE}

If the continuous extension of physical space and the spatial multiplicity of 
the physical world do not exist by courtesy of the mind/brain system, then 
how shall we conceive of them? The answer, in brief, is 
this~\cite{UMCCP,BCCP}: Physical space is the totality of spatial relations, 
or relative positions, that exist between material objects. The spatial 
multiplicity of the physical world, accordingly, consists in the (discrete) 
multiplicity of existing spatial relations. These relations have a qualitative as 
well as a quantitative aspect. The former consists in the quality of spatial 
extension; it constitutes the spatial character of each relation and does not 
connote any kind of multiplicity. The phenomenal world owes its spatial 
extension to the mental canvas on which it appears. If we accord a 
mind-independent reality to the physical world, we cannot attribute its spatial 
extension to an independently existing expanse on which material things 
are ``displayed.'' Instead we must attribute it to each spatial relation. The 
spatial extension of the physical world is a quality that all spatial relations 
share.

It is worth noting, in passing, that the concept of an intrinsically partitioned 
physical space makes it impossible to understand the spatial unity of the 
physical world---except in weakly objective terms, as the projected unity of 
our mental canvas. Assume that physical space is a thing with parts and 
you are confronted with the task of explaining what holds together the parts 
of space. To my way of thinking, the proper way to cope with this task is not 
to take refuge in the secure haven of weak objectivity but to acknowledge 
the neural basis of the special treatment we accord to locations, to restrict 
this special treatment to the phenomenal world, and to treat physical 
locations in the same way as we treat the color of a ripe tomato or the 
shape of a sphere---as existing only if possessed. The task then reduces 
itself to explaining what holds together the parts of a material object. The 
purpose of this paragraph was to demonstrate that the concept of an 
intrinsically partitioned physical space is all that is needed to forestall any 
strongly objective conception of the physical world. This, however, should 
prompt us to look for a different concept of physical space, rather than to 
the rejection of strong objectivity.

There is mounting evidence from neuroscience that visual perception and 
visual imagination share the same processing 
mechanisms~\cite{Finke,ShepardCooper}. It is therefore to be expected 
that the inherent graininess of phenomenal space also conditions our 
non-sensory visual images, including the images we form of hydrogen orbitals. 
Once we realize that the graininess originates in the neural constitution of 
our feature maps, we should find it easier to see that the parts of such an 
image do not represent material parts. Where there are material parts there 
are as many positions as there are material parts. Since the electron cloud 
represents a single (relative) position, such an image represents a single 
partless object. The parts of the image have a {\it counterfactual\/} reality, 
as was explained in Sec.~\ref{TSE}. What about the {\it extension} of the 
electron cloud? Remarkably, this {\it does} represent an actual property of 
the hydrogen atom's internal spatial relation, namely its quality of being 
spatial or extended. This shows that physical space, qua extension, can 
exist where spatial parts do not exist.

\section{\large OBJECTIVE INDEFINITENESS AND THE FACTUALITY 
PROBLEM}

It is well known that the objective indefiniteness of relative positions is 
crucial for the existence of stable, extended material objects~\cite{Lieb}. 
Together with the exclusion principle this is what ``fluffs out'' matter. Hence 
it would seem that finding a satisfactory formal expression of objective 
indefiniteness should be of paramount importance. The proper way of 
dealing with objectively indefinite values is to make counterfactual 
probability assignments~\cite{Mohrhoff00,UMABL,UMWATQM}. If a 
quantity is said to have an ``indefinite value,'' what is really intended is that 
it does not actually have a value (inasmuch as the value is not measured) 
but that it {\it would} have a value if this {\it were} indicated, and that at least 
two possible values are associated with positive probabilities.

Since an observable may or may not have a value, we need a criterion for 
deciding when it has a value, and this criterion consists in the existence of 
a value-indicating fact. The extrinsic nature of the values of 
quantum-mechanical observables (Sec.~\ref{TSE})---a rephrasing of the well-known 
necessity of describing quantum phenomena in terms of the experimental 
arrangements in which they are displayed~\cite{BohrEssays}---is therefore 
a straightforward consequence of their objective indefiniteness. Objective 
indefiniteness entails the possible lack of a value, which entails the need for 
a criterion, which consists in the existence of a value-indicating fact.

The problem of understanding QM is sometimes portrayed as the problem 
of explaining the emergence of facts 
(``classicality'')~\cite{JoosZeh,Zurek91,Zurek93}, but this cannot be taken 
literally. QM is concerned with correlations between property-indicating 
facts---diachronic correlations between results of measurements performed 
on the same system and synchronic correlations between results of 
measurements performed on entangled systems. QM presupposes facts, 
and therefore it cannot account for their existence. Nor can any other 
theory. Accounting for the existence of facts is the same as explaining why 
there is anything at all, rather than nothing---an impossible task.

In this regard quantum physics is no different from classical physics. 
Classical physics deals with nomologically possible worlds (worlds 
consistent with physical theory). It does not tell us which of all possible 
worlds corresponds to the actual world. Much the same is true of quantum 
physics. The main difference is that in classical physics the actual course 
of events is in principle fully determined by the actual initial conditions (or 
the actual initial and final conditions), while in quantum physics it also 
depends on unpredictable facts at later (or intermediate) times. There are 
just many more possible worlds, and there are many more extra-theoretical 
conditions to be satisfied by a possible world if it is to correspond to the 
actual world.

In order to solve the solid core of the measurement problem one must show 
that certain properties can be regarded as intrinsic---as factual {\it per 
se}---not only for all practical purposes but for all {\it quantitative} 
ones~\cite{Mohrhoff00,UMCCP}. The proof proceeds in three steps.

(1)~Since no position is possessed unless it is indicated, and since nothing 
ever indicates an exact position, nothing ever has an exact position. Hence 
there exists some finite limit to the sharpness of the positions of material 
objects, and there exists some finite limit to the spatial resolution of actual 
detectors. Consequently there are objects that never evince their positional 
indefiniteness through unpredictable position-indicating 
facts---objects that evolve predictably in the sense that every time the 
position of such an object is indicated, its value is consistent with all 
predictions that can be made on the basis of (i)~past indicated properties 
and (ii)~classical laws of motion. This follows from the fact that evidence of 
departures from the pertinent classical laws requires detectors with 
sensitive regions that are small and localized enough to probe the range of 
values over which a position is distributed. If there is a finite limit to the 
sharpness of the positions of material objects, there are objects that have 
the sharpest positions in existence. The positions of such objects, which 
deserve to be called ``macroscopic,'' cannot but evolve predictably. We 
cannot be certain that a given object qualifies as macroscopic, inasmuch as 
not all relevant position-indicating facts are accessible to us, but we can be 
certain that macroscopic objects exist.

(2)~If the positional indefiniteness of a macroscopic object never evinces 
itself through unpredictable position-indicating facts---the occasional 
unpredictability of the position of a macroscopic pointer is an indication of 
the indefiniteness of a different 
observable---then what kind of reality does the positional indefiniteness of a 
macroscopic object possess? If we make the assumption that macroscopic 
objects follow definite trajectories, we will never see this assumption 
contradicted by facts. If instead we think of the position of a macroscopic 
object---a ``macroscopic position'' for short---as a ``wave packet,'' this is 
distributed over regions of space that are never objectively distinct. Such 
regions exist only in our imagination. They are the sensitive regions of 
detectors that do not exist in the physical world. They represent an 
unrealized degree of spatial differentiation. But if the position of an object is 
distributed over regions that do not exist, it is not {\it actually} distributed. 
The indefiniteness of a macroscopic position therefore has never more than 
a {\it counterfactual\/} reality.

(3)~Now recall that the objective indefiniteness of the values of 
quantum-mechanical observables is the reason why they are extrinsic. Since the 
indefiniteness of a macroscopic position never evinces itself in the realm of 
facts, we can think of macroscopic positions as forming a self-contained 
system of positions that ``dangle'' causally from each other, rather than 
ontologically from position-indicating facts. We can ignore their extrinsic 
nature, consider them as intrinsic, and thus as factual {\it per se}. We can 
then attribute the possession of value, by a quantum-mechanical 
observable, to the value's being indicated by at least one macroscopic 
position.

\section{\large CONCLUSIONS}

The transition from positions that supervene on facts to positions that form 
a 
self-contained causal nexus is quantitatively impeccable, inasmuch as the 
statistical correlations between indicated macroscopic positions 
are completely dispersion-free. Conceptually the transition is of the same 
nature as the transition from a purely correlative interpretation of the 
classical laws of motion (that is, from causality qua {\it regularity}) to an {\it 
efficient} interpretation that posits causal links responsible for the regularity. 
The latter transition is at bottom nothing but the projection, into the 
time-symmetric world of classical physics, of our own time-asymmetric agent 
causality~\cite{Mohrhoff00,UMWATQM}.

Classical physics is concerned with deterministic correlations that admit of 
a causal interpretation; quantum physics is concerned with probabilistic 
correlations that don't, except in the classical domain. This is where the 
quest for meaning really begins, not where it ends. For the probabilistic 
correlations are trying to tell us something, something that concerns the 
nature of physical space.

While the whereabouts of macroscopic objects are abundantly indicated, 
they are never indicated with absolute precision. Hence even for 
macroscopic objects the world at any given time $t$ is only finitely 
differentiated spacewise. That is, no finite region $R$ is differentiated into 
infinitely many regions $R_i$ such that truth values exist for all propositions 
of the form ``$O$~is inside $R_i$ at the time~$t$,'' where $O$ is any 
macroscopic object. And this---the {\it finite spatial differentiation of the 
physical world\/}---is arguably the single most significant ontological 
implication of QM~\cite{Mohrhoff00,UMCCP}. The world is created 
top-down, by a finite process of differentiation, rather than built bottom-up, on 
an infinitely differentiated space, out of locally instantiated physical 
properties. The field-theoretic notion of an intrinsically and infinitely 
differentiated space is therefore as inconsistent with QM as the notion of 
absolute simultaneity is with special relativity. Spatial distinctions are not 
intrinsic to space. They supervene on the facts. These constitute a domain 
whose indefiniteness exists only in relation to an unrealized degree of 
spatial differentiation (that is, only in relation to an imaginary backdrop that 
is more differentiated spacewise than is the physical world). Atoms exist 
beyond the realized degree of spatial differentiation, and this is what makes 
them so decidedly\dots {\it alien}.

To recapitulate, the indefiniteness of relative positions contributes to ``fluff 
out'' matter. The indefiniteness of the values of quantum-mechanical 
observables entails their extrinsic nature: No value is a possessed value 
unless it is an indicated value. This implies the contingent reality of spatial 
distinctions, and combined with a dynamical equation such as the 
Schr\"odinger equation, it implies the finite spatial differentiation of the 
physical world. This makes it possible to rigorously distinguish between the 
classical and the quantum domains, and to understand their mutual 
dependence. While macroscopic objects, like all composite objects, owe 
their finite volumes to the indefiniteness of their internal spatial 
relations and the exclusion principle, the properties of the quantum domain 
owe their existence to the 
classical domain: A position-indicating fact is necessary for the possession 
of a position, and a macroscopic detector is necessary for the existence of 
an attributable position---it realizes this position, counterfactually if not in 
actual fact.

The solution of a major puzzle posed by QM thus stands and falls with the 
finite spatial differentiation of the physical world. It cannot be found unless 
we relinquish our psychologically and neurobiologically sustained belief in 
an intrinsically (and therefore infinitely) differentiated space. It is this 
deep-seated but physically unwarranted belief that is responsible for our failure, 
so far, to find a sensible ontological interpretation of QM, and for our 
tendency to take refuge in epistemic interpretations.

\appendix\noindent Abstract (German translation): Epistemische Deutungen der 
Quantenmechanik l\"osen nicht das R\"atsel, vor das uns das Vorkommen von 
Wahrscheinlichkeiten in einer grundlegenden phy\-si\-ka\-li\-schen Theorie stellt. 
Dieses R\"atsel betrifft nicht unsere Beziehung zur physischen Welt, sondern 
die physische Welt selbst. Seine L\"osung erfordert einen neuen Raumbegriff, 
der in diesem Artikel vorgestellt wird. Eine Untersuchung der geistigen und 
neurobiologischen Wurzeln der Erscheinungswelt kl\"art, warum unser Denken 
\"uber den Raum der physischen Welt nicht gerecht wird. Die diesem Denken 
zugrunde liegende Idee, dass der Raum innerlich zerteilt ist, stand einer 
konsequenten ontologischen Deutung bisher im Wege.

\end{document}